\documentclass[11pt]{JHEP3}

\JHEPspecialurl{http://jhep.sissa.it/JOURNAL/JHEP3.tar.gz}

\usepackage{epsfig,multicol}
\usepackage{amsmath,amssymb}
\usepackage{mathrsfs}

\newcommand\fverb{\setbox\fverbbox=\hbox\bgroup\verb}
\newcommand\fverbdo{\egroup\medskip\noindent%
			\fbox{\unhbox\fverbbox}\ }
\newcommand\fverbit{\egroup\item[\fbox{\unhbox\fverbbox}]}
\newbox\fverbbox

\newcommand{\pslash}{p\kern-1ex /}
\newcommand{\qslash}{q\kern-1ex /}
\newcommand{\lslash}{l\kern-1ex /}
\newcommand{\sslash}{s\kern-1ex /}
\newcommand{\kaslash}{k_a\kern-2ex /}
\newcommand{\kbslash}{k_b\kern-2ex /}
\newcommand{\Dslash}{\mathcal{D}\kern-1.5ex /}

\newcommand{\beqa}{\begin{eqnarray}}
\newcommand{\eeqa}{\end{eqnarray}}

\newcommand{\ba}{\begin{eqnarray}}
\newcommand{\ea}{\end{eqnarray}}
\newcommand{\be}{\begin{equation}}

\title{The Bajnok-Janik formula and wrapping corrections}

\author{J\'anos Balog and \'Arp\'ad Heged\H us\\
Research Institute for Particle and Nuclear Physics, 
Hungarian Academy of Sciences,
H-1525 Budapest 114, P.O.B. 49, Hungary\\}

\received{\today} 		
\accepted{\today}	

\abstract{
We write down the simplified TBA equations of the $AdS_5 \times S^5$ 
string $\sigma$-model for minimal energy twist-two 
operators in the $sl(2)$ sector of the model.
By using the linearized version of these TBA equations it is shown that the
wrapping corrected Bethe equations for these states  are identical, 
up to $O(g^8)$, to the Bethe equations calculated in the generalized 
L\"uscher approach (Bajnok-Janik formula).
Applications of the Bajnok-Janik formula to relativistic integrable
models, the nonlinear O$(n)$ sigma models for $n=2,3,4$ and the SU$(n)$
principal sigma models, are also discussed.
}

\begin{document}

\section{Introduction}

One of the most important problems in testing the AdS/CFT correspondence 
\cite{adscft} is to understand
the finite size spectrum of the $AdS_5 \times S^5$ superstring. For large 
volumes the asymptotic
Bethe Ansatz (ABA) describes the spectrum of the model \cite{BS}. 
It takes into account all power like
corrections in the size, but neglects the exponentially small wrapping corrections \cite{AJK}.

In \cite{BJ08,JL07} it was shown that the leading order wrapping corrections can also be expressed by the 
infinite
volume scattering data through the generalized L\"uscher formulae \cite{Luscher85}.
In \cite{BJ08} the 4-loop anomalous dimension of the Konishi operator was obtained by means of the generalized
L\"uscher formulae in perfect agreement with direct field theoretic computations \cite{Sieg,Vel}.
Subsequently wrapping interactions computed from L\"uscher corrections were found to be crucial for the
agreement of some structural properties of twist-two operators \cite{Bajnok:2008qj} with LO and NLO BFKL 
expectations \cite{KL02,40_5}.

More recently \cite{BJ09} the 5-loop wrapping correction to the anomalous dimension of the Konishi operator was 
also computed from the generalized L\"uscher approach accounting for the expected nontrivial transcendentality
structure of the anomalous dimension.
%
Later the 5-loop result has been extended to the class of twist two operators as well
\cite{Lukowski:2009ce}. After analytic continuation to negative values of the spin this
gave nontrivial agreement with the predictions of the BFKL equations \cite{KL02}.

Although the generalized L\"uscher approach was invented for the purpose 
of computing the wrapping corrections in the AdS/CFT context, it has more 
general validity. In particular, it is also valid in relativistic integrable
models, like the Sine-Gordon model, nonlinear $\sigma$-models and other related
models. Based on the TBA/NLIE description of these models, we have proven the
validity of the Bajnok-Janik approach to generalized L\"uscher corrections
for them. In appendix C we summarize the results for the O$(2)$, O$(3)$ and O$(4)$ 
nonlinear $\sigma$-models and the SU$(n)$ principal model.

After the discovery of integrability of the string worldsheet theory the mirror 
Thermodynamic Bethe Ansatz (TBA) technique was used \cite{AJK,AF07} to determine the exact spectrum of string 
theory (including the exponentially small L\"uscher corrections). The TBA 
equations of AdS/CFT were derived first for the ground state \cite{AF09a,AF09b,AF09d,Bombardelli:2009ns,GKKV09} 
and then using an analytic continuation trick \cite{DT} excited states TBA equations were conjectured for the
excitations of the $sl(2)$ sector of the theory \cite{GKKV09,GKV09b,Arutyunov:2009ax}.
Since the final form of the TBA equations is still a conjecture it is important to test them carefully.  
In the strong coupling limit it was shown \cite{Gromov,ujKazi} that the TBA equations reproduce 
correctly the 1-loop
string energies in the quasi-classical limit. On the other hand in the opposite weak coupling limit it is
of fundamental importance to see that the TBA equations are consistent with the generalized L\"uscher formulae.
This can be tested by studying the small $g$ expansion{\footnote{Here $g$ is the coupling constant 
related to the 't Hooft coupling $\lambda$ through $\lambda=4\pi^2g^2$.}} of both the L\"uscher formulae 
and the TBA equations.

In the TBA context the string energies are given by the formula:
\begin{equation} \label{E}
E=J+\sum_{i=1}^N{\cal
E}(p_i)-\frac{1}{2\pi}\sum_{Q=1}^{\infty}\int_{-\infty}^{\infty}{\rm
d }u\frac{d\widetilde{p}^Q}{du}\log(1+Y_Q)\,,  
\end{equation} 
where $N$ is the number of particles, $J$ is
the angular momentum carried by the string rotating around the
equator of $S^5$, $\widetilde{p}^Q$ is the mirror momentum
and the functions $Y_Q$ are the unknown functions (Y-functions) associated to the mirror $Q$-particles,
furthermore 
\begin{equation} {\cal E}(p)=\sqrt{1+4g^2\sin^2\frac{p}{2}}
\end{equation}
is the dispersion relation of the string theory particles.

The wrapping corrections to string energies have two sources: the momenta of the particles change due to
wrapping corrections to the ABA and there is a contribution proportional to the asymptotic form of the $Y_Q$
functions. Since the asymptotic form of the $Y_Q$ functions are built into the TBA equations by construction,
only the consistency of the wrapping corrected forms of the ABA obtained from the 
TBA and the generalized L\"uscher approach has to be verified.
This comparison has been done first numerically \cite{AFS} then
analytically \cite{BHL} for the Konishi state in leading 
order{\footnote{I.e. at the order of $g^8$.}} in $g$. 
In this paper we extend this comparison to the familiy of 
minimal energy twist-two operators $\mbox{Tr}(D^N Z^2)+\dots$  of the 
$sl(2)$ sector of the theory.

According to the generalized L\"uscher approach the leading order wrapping corrected Bethe equations
for the twist-two ($J=2$) operators take the form:
\begin{eqnarray} \label{WCABA}
\pi (2n_k+1)&=&J\,
p_k +i\sum_{j=1}^N \log S_{\mathfrak{sl}(2)}^{1_*1_*}({u_{j}}{},{u_{k}}{}) +{\cal
\delta R}_k^{({\rm BJ})}\,+O(g^9), 
\end{eqnarray}
where $n_k$ is the integer quantum number characterizing the corresponding rapidity $u_k$ and
$\delta{\cal R}_k^{({\rm BJ})}$ is the order $g^8$ wrapping correction to the ABA, obtained from 
the Bajnok-Janik formula \cite{BJ08}: 
\begin{equation}
\delta{\cal R}_k^{({\rm BJ})}=
\frac{1}{2\pi}\,
\sum_{Q=1}^{\infty}\int_{-\infty}^{\infty}{\rm d}u\, \,
\frac{\partial}{\partial u_k} \, Y_Q^{asympt}(u)\big|_{ \{u_j\}=\{u^o_j\} }   \,.
\label{BJv0}
\end{equation}
Here $Y_Q^{asympt}(u)$ is of order $g^8$ and is a function of $u$ and the particle 
rapidities $u_j$. After the differentiation the particle rapidities are taken at the 
solution of the ABA, $u^o_j$. We note that the simple formula (\ref{BJv0})
is valid only for the leading order wrapping corrections of the Bethe equations. At higher orders
in $g$ the corrections can no longer be expressed by the derivative of the $Y_Q$ functions with respect to 
the magnon rapidities $u_k$. The more general nonrelativistic formula can be found in \cite{BJ08}.

In this paper we will prove that by expanding the TBA equations we can exactly reproduce the formulae 
(\ref{WCABA}),(\ref{BJv0})
for the $g^8$ order wrapping corrections of the Bethe equations for the twist-two operators.
The proof is a generalization of that used for the case of the Konishi operator in \cite{BHL}.

The paper is organized as follows. In section 2 we write down the TBA 
equations for the twist-two operators. In section 3 we linearize them around 
the asymptotic solution to describe the wrapping  effects. In section 4 we 
present our results and the paper is finished with some conclusions.
The technical details of the calculation are discussed in appendix~A and
we give a derivation of the leading order Bajnok-Janik formula (\ref{BJv0}) 
in appendix~B. Appendix~C contains the Bajnok-Janik formula for some
relativistic integrable models.

\section{Simplified TBA equations for the twist-two operators}

Excited state TBA equations were proposed for certain classes of states in the $sl(2)$ 
sector in \cite{GKKV09,Arutyunov:2009ax}. The TBA equations of 
ref.~\cite{GKKV09} are valid for states  
where only the singularities associated to
the function $Y_1$ have to be taken into account while 
\cite{Arutyunov:2009ax} contains the detailed analysis of the two-particle 
states. Since our states of 
interest are not discussed{\footnote{apart from the Konishi operator}} in the above papers 
here we write down the TBA equations for the twist-two operators valid for small 
values of the coupling. We proceed in the spirit of \cite{Arutyunov:2009ax}, namely it is assumed that
the TBA equations are formally the same for the ground state and excited states provided the integration 
contours are defined properly. The form of the equations becomes different when the integration
contours are deformed back to the real line of the mirror theory picking up the contribution of the 
poles from the convolution terms.
The necessary singularity structure of the Y-functions can be read off from their asymptotic form. 

The twist-two operators are given by $N/2$ pair{\footnote{$N$ is even}} 
of real rapidities $\{u_j,-u_j\}$ and the
asymptotic form of the Y-functions associated to these states can be 
obtained by using the results of \cite{GKV09}.
Inspecting the analyticity properties of the Y-functions the TBA equations for the twist-two operators can be
written down and take the following form{\footnote{Here we use the conventions, terms and 
notations of ref. \cite{Arutyunov:2009ax}}}:

\bigskip
 \noindent
$\bullet$\ $M|w$-strings: $\ M\ge 1\ $, $Y_{0|w}=0$
\begin{eqnarray} \label{YM|w}
\log Y_{M|w}=  \log(1 +  Y_{M-1|w})(1 +  Y_{M+1|w})\star s
+\delta_{M1}\, \log{1-{1\over Y_-}\over 1-{1\over Y_+} }\,\hat{\star}\, s\,.~~~~~
\end{eqnarray}

\bigskip
 \noindent
$\bullet$\ $M|vw$-strings: $\ M\ge 1\ $, $Y_{0|vw}=0$
\begin{eqnarray} \label{YM|vw}
&&\hspace{-0.3cm}\log Y_{M|vw}(v)=-\delta_{M1} \sum_{j=1}^N \log S(u_j^--v)
-\sum\limits_{M'=1}^{\infty} I_{M,M'} \sum_{j=1}^{n_{M'}} \log S(r_j^{(M')-}-v) ~~~~~\\\nonumber
&&+ \log(1 +  Y_{M-1|vw} )(1 +  Y_{M+1|vw})\star s+\delta_{M1}  \log{1-Y_-\over 1-Y_+}\,\hat{\star}\, s- \log(1 
+  Y_{M+1})\star s\,,~~~~~
\end{eqnarray}
with $I_{M,M'}=\delta_{M,M'-1}+\delta_{M,M'+1}$ and $n_M=2(N-2)$ if $M\geq 1$, $n_0=0$.
The first term is due to the pole of $Y_+$ at $u=u_j^-=u_j-\frac{i}{g}$, and  the second term is due to the  
zeros of $1 + Y_{M|vw}$ at $u=r^{(M)-}_j=r^{(M)}_j-\frac{i}{g}$ which are subject to the 
quantization conditions:
\begin{equation}
\log Y_{M|vw}(r^{(M)-}_j)=2 \pi \, i \, I^{(M)}_{j}, \qquad r^{(M)}_j \in \mathbb{R},
\end{equation}
where the $I^{(M)}_{j}$s are half-integer quantum numbers. 

\bigskip
 \noindent
$\bullet$\   $y$-particles:
\begin{eqnarray}\label{Y+/-}
\log {Y_+\over Y_-}(v)&=& -\sum_{j=1}^N \log S_{1_*y}(u_{j},v)  +\log(1 +  Y_{Q})\star K_{Qy}\,,~~~~~~~
 \\
\label{Y+-}
\log {Y_+ Y_-}(v) &=& -\sum_{j=1}^N\,
 \log {\big(S_{xv}^{1_*1}\big)^2\over S_2}\star s(u_j,v)-2\sum_{j=1}^{n_1} \log S(r^{(1)-}_j-v)
\\\nonumber
&+&2\log{1 +  Y_{1|vw} \over 1 +  Y_{1|w} }\star s
- \log\left(1+Y_Q \right)\star K_Q+ 2 \log(1 +  Y_{Q})\star K_{xv}^{Q1} \star s\,,~~~~
\end{eqnarray}
where the second term in the second line  is due to the  zeros of $1 +  Y_{1|vw}$ at $u=r^{(1)-}_j$.

\bigskip
 \noindent
$\bullet$\  $Q$-particles for $Q\geq 2$
\begin{eqnarray}
\log Y_{Q}&=& - 2\sum_{j=1}^{n_{Q-1}} \log S(r_{j}^{(Q-1)-}  -v)+
\log{\left(1 +  {1\over Y_{Q-1|vw}} \right)^2\over (1 +  {1\over Y_{Q-1} })(1 +  {1\over Y_{Q+1} }) 
}\star_{p.v.} s\label{YQ2}\,.
\,~~~~~~~
\end{eqnarray}
where the source term on the right hand side comes from the zeroes of $1 +  Y_{Q-1|vw}$ at $u=r^{(Q-1)-}_j$.
We note that the p.v. prescription is only necessary for $Q=2$.
For the $Q=1$ case the hybrid version \cite{Arutyunov:2009ax} of the TBA equations is 
more useful.
\bigskip

 \noindent
$\bullet$\  Hybrid equation for the $Q=1$ particle
\begin{align}
&\log Y_1(v) = - \sum_{j=1}^N (  \log S_{\mathfrak{sl}(2)}^{1_*1}(u_j,v)
 - 2 \log S\star K^{11}_{vwx} (u_j^-,v) )
\notag\\
&\quad - L\, \tilde{\cal E}_{1}
+ \log \left(1+Y_{Q'} \right) \, \star \, (K_{\mathfrak{sl}(2)}^{Q'1} + 2 \, s \, \star \, K^{Q'-1,1}_{vwx} )
\label{HY1} \\
&\quad 
-2 \sum\limits_{j=1}^{n_1} \left(  \log S \, \hat{\star} \, K_{y1} \right) (r_j^{(1)}-\frac{i}{g},v)
+ 2 \log (1 + Y_{1|vw}) \, \star \, s \, \hat{\star} \, K_{y1}
\notag \\
&\quad - 2  \log{1-Y_-\over 1-Y_+} \, \hat{\star} \, s \, \star \, K^{11}_{vwx}
+  \log {1- \frac{1}{Y_-} \over 1- \frac{1}{Y_+} } \, \hat{\star} \, K_{1} +  \log \big(1- 
\frac{1}{Y_-}\big)\big( 1- \frac{1}{Y_+} \big) \, \hat{\star} \, K_{y1} \,,
\notag
\end{align} 
where the first term on the right hand side comes from the zeros of $1+Y_1$ at the magnon rapidities $u_j$
while the source term in the third line of (\ref{HY1}) is due to the  zeros of $1 +  Y_{1|vw}$ at 
$u=r^{(1)-}_j$.
Here $L=2+J$ with $J=2$ for the twist-two operators. We believe that this
relation between the charge $J$ and length $L$ is valid for all states
in the $sl(2)$ sector with a symmetric distribution of magnon rapidities.
These include the minimal energy twist-two states discussed in this 
paper and all $N=2$ states studied in \cite{Arutyunov:2009ax}.


The analytical continuation of (\ref{HY1}) yields the exact Bethe equations for the real rapidities:
\begin{align}
&\pi i(2n_k+1)=\log Y_{1_*}(u_k) =i L\, p_k- \sum_{j=1}^N\, \log S_{\mathfrak{sl}(2)}^{1_*1_*}(u_j,u_k)
\label{Y1*}\\
&\quad
 + 2 \sum_{j=1}^N\, \log {\rm Res}(S)\star K^{11_*}_{vwx} (u_j^-,u_k) -2 \sum_{j=1}^N\log\big(u_j-u_k-{2i\over 
g}\big)\,
{x_j^--{1\over x_{k}^-}\over x_j^-- x_{k}^+}
\notag\\
&\qquad\qquad\quad  - 2\sum_{j=1}^{n_1} \left( \log S\,\hat{\star}\, K_{y1_*}(r_j^{(1)-},u_k)- \log 
S(r^{(1)}_j-u_k)\right) \notag\\
&\quad
+ \log \left(1+Y_{Q} \right) \star \left(K_{\mathfrak{sl}(2)}^{Q1_*} + 2 \, s \star K^{Q-1,1_*}_{vwx} \right)+ 
2 \log \left(1 + Y_{1|vw}\right) \star \left( s \,\hat{\star}\, K_{y1_*} + \tilde{s}\right)
\notag \\
&\quad - 2  \log{1-Y_-\over 1-Y_+} \,\hat{\star}\, s \star K^{11_*}_{vwx}
+  \log {1- \frac{1}{Y_-} \over 1- \frac{1}{Y_+} } \,\hat{\star}\, K_{1} +  \log \big(1- 
\frac{1}{Y_-}\big)\big( 1- \frac{1}{Y_+} \big) \,\hat{\star}\, K_{y1_*} \,.
\notag
\end{align}
The source and kernel functions together with the definition of the convolutions $\star$ and $\hat{\star}$ 
appearing in (\ref{YM|w})-(\ref{Y1*}) can be found in \cite{Arutyunov:2009ax}.

\section{The linearized problem}

The $O(g^8)$ wrapping correction{\footnote{In the case of the 
twist-two ($J=2$) operators
the fact that the wrapping corrections start only at $O(g^8)$ can 
also be understood
using superconformal invariance \cite{BS,SCFT1}.}} 
to the ABA can be expressed by perturbing the TBA equations 
(\ref{YM|w}-\ref{Y1*}) around the asymptotic solution \cite{AFS}. 
Borrowing the notation from \cite{AFS} for any $Y$ function let $Y^{o}$ be its asymptotic expression and 
${\mathscr Y}$ the exponentially small perturbation around $Y^{o}$ defined by the equation:
\begin{eqnarray}
Y=Y^{o}(1+{\mathscr Y}).  
\label{mathscrY}
\end{eqnarray}
Similarly we define the perturbation of the zeroes of $1+Y_{M|vw}$ by the formula:
\begin{equation}
r^{(M)}_j=\hat{r}^{(M)}_j+\delta r^{(M)}_j,
\end{equation}
with $\hat{r}^{(M)}_j$ being the asymptotic value and $\delta r^{(M)}_j$ the small perturbation.
Strictly speaking ${\mathscr Y}_{M|vw}$, as defined by (\ref{mathscrY}) cannot be small everywhere, since
the zeroes of the asymptotic solution are shifted by the perturbation and therefore ${\mathscr Y}_{M|vw}$
appears to have poles on the real axis. This problem can be avoided if we shift the integration contour
for the $M|vw$ equations by (a small amount) $i\gamma$ below the real line before linearization. It is 
easy to see that the shifted contour does not cross any singularities and in the formulas (\ref{dYM|vw})
and (\ref{ADSlin}) below this shift is understood (but not indicated explicitly). The actual calculation
is performed in appendix A where the shifted contour is used throughout. 

Expanding the TBA equations around the asymptotic values one finds the following set of linear
equations for the perturbations:

\bigskip
 \noindent
$\bullet$\ $M|w$-strings: $\ M\ge 1\ $, ${\mathscr Y}_{0|w}=0$
\begin{eqnarray}\label{dYM|w} 
{\mathscr Y}_{M|w}=  (A_{M-1|w}{\mathscr
Y}_{M-1|w}+ A_{M+1|w}{\mathscr Y}_{M+1|w})\star s +\delta_{M1}\,
\Big(\frac{{\mathscr Y}_+}{1-Y_+^o}-\frac{{\mathscr
Y}_-}{1-Y_-^o}\Big)\, \hat{\star} \, s\, ,~~~~~ 
\end{eqnarray} 
where  $A_{M|w}=\frac{Y^o_{M|w}}{1+Y^o_{M|w}}$. 

\bigskip
 \noindent
$\bullet$\ $M|vw$-strings: $\ M\ge 1\ $, ${\mathscr Y}_{0|vw}=0$
\begin{eqnarray}\label{dYM|vw} &&{\mathscr Y}_{M|vw}=  (A_{M-1|vw}{\mathscr
Y}_{M-1|vw}+ A_{M+1|vw}{\mathscr Y}_{M+1|vw})\star s
-Y_{M+1}^o\star s \\
\nonumber 
&&\hspace{1cm}
-2 \pi i \sum\limits_{M'=1}^{\infty} \, I_{M,M'} \, \sum\limits_{j=1}^{n_{M'}}
s(\hat{r}^{(M')}_j-\frac{i}{g}-v) \, \delta r^{(M')}_j
+\, \delta_{M1}\, \Big(\frac{{\mathscr
Y}_-}{1-\frac{1}{Y_-^o}}-\frac{{\mathscr Y}_+}{1-\frac{1}{Y_+^o}}
\Big)\, \hat{\star} \, s\, \,,~~~~~ \end{eqnarray}
with $A_{M|vw}\equiv {Y_{M|vw}^o\over 1+Y_{M|vw}^o }$. There are additional linear equations for the
perturbations of the $r^{(M)}_j$s:
\begin{equation}\label{drMj}
(\log Y^{o}_{M|vw})'(\hat{r}^{(M)}_j-\frac{i}{g}) \, \delta r^{(M)}_j+{\mathscr 
Y}_{M|vw}(\hat{r}^{(M)}_j-\frac{i}{g})=0, \qquad  M \geq 1, \qquad j=1,...,n_M 
\end{equation}
where the prime means differentiation with respect to the argument.

\bigskip
 \noindent
$\bullet$\   $y$-particles \begin{eqnarray}\label{dY+/-} &&{\mathscr Y}_+ -
{\mathscr Y}_-= Y_{Q}^o\star K_{Qy}\,,~~~~~~~ \\
\label{dY+-}
 &&{\mathscr Y}_+ +
{\mathscr Y}_-= 2(A_{1|vw}{\mathscr Y}_{1|vw}-A_{1|w}{\mathscr
Y}_{1|w})\star s -4 \pi i \sum\limits_{j=1}^{n_1} \, s(\hat{r}^{(1)}_j-\frac{i}{g}-v)
\, \delta r^{(1)}_j \nonumber \\
&& -Y_Q^o\star s+2 Y_Q^o\star K_{xv}^{Q1}\star s\,
. \end{eqnarray}

Now we are in a position to analyze the magnitudes of the different terms with respect to the coupling $g$.
For this purpose we further expand eqs. (\ref{dYM|w})-(\ref{dY+-}) with respect to the coupling.
The source terms of the linear equations for the perturbations are given by convolution terms of the 
$Y^{o}_Q(u)$ functions. The $Y^{o}_Q(\frac{u}{g})$ functions can be generated using the results of ref. 
\cite{GKV09}, and it turns out that similarly to the case of the Konishi field they are of $O(g^8)$ for small 
$g$ 
\cite{Bajnok:2008qj}. Using this fact in can be shown that the 
convolution terms of the linear problem containing the  
$Y^{o}_Q(u)$ functions are at least of order $g^8$. This implies that also the perturbations are at least
of $O(g^8)$. As $Y_Q^o\star K_{xv}^{Q1}\star s= O(g^8)$, from (\ref{dY+-}) 
it follows that 
${\mathscr Y}_{\pm} =O(g^8)$, while (\ref{dY+/-}) implies 
that ${\mathscr Y}_{+} -{\mathscr Y}_{-}=O(g^9)$ 
since $Y_{Q}^o\star K_{Qy}=O(g^9)$. 

Further from the asymptotic solution of the $Y$-functions we see that  
$Y^{o}_{+}$ and $Y^{o}_{-}$ coincide at leading order in $g$:
$$
\frac{Y^{o}_{+}(u)}{Y^{o}_{-}(u)}=1+O(g^2).
$$ 
This implies according to (\ref{dYM|w}) that ${\mathscr Y}_{M|w}(\frac{u}{g})=O(g^9)$ and that the order 
$g^8$ perturbations of the $Y_{M|vw}$s in (\ref{dYM|vw}),(\ref{drMj}) are unaffected by the contributions of 
the 
$Y_{\pm}$ functions and decouple from the other type of variables.

\section{The Bajnok-Janik formula}

Finally we can turn our attention to the Bethe equations (\ref{Y1*}). In (\ref{Y1*}) the 
$Y_{\pm}$ functions appear only through $\hat{\star}$ type convolution terms this is why their perturbations
give only $O(g^9)$ contribution. Consequently up to the order of $g^8$ only the
perturbations of the $Y_{M|vw}$ functions contribute. 
Rescaling the variables $u\rightarrow u/g$, $u_k\rightarrow u_k/g$,
 $\hat{r}^{(m)}_j\rightarrow \xi_{m;j}/g$ and $\delta {r}^{(m)}_j\rightarrow \delta \xi_{m;j}/g$ and
making similar considerations as in \cite{AFS} and \cite{BHL}
it can be shown that the Bethe equations (\ref{Y1*}) up to $O(g^8)$ can be expressed by
the leading $O(g^8)$ expressions of the $Y^{o}_Q$ functions and with the solution of a linear problem
coming from (\ref{dYM|vw}). This linear problem is similar to the linearization of the TBA equations 
of the XXX Heisenberg chain \cite{BHL} and in the rescaled variables it takes the form:
\begin{equation} \label{ADSlin}
\frac{\delta y_m}{y_m} -s \star (\delta L_{m-1}+\delta L_{m+1})+
\sum\limits_{m'=1}^{\infty} I_{m,m'} \, 
\sum\limits_{j=1}^{n_{m'}} g(u-\xi_{m';j})\delta\xi_{m';j}
=-s \star Y^{o}_{m+1}, \quad
m=1,2,...,
\end{equation}
\begin{equation} \label{ADSlinquant}
\left(\frac{\delta y_m}{y_m} \right)(\xi_{m;j}-i)-y^{\, 
\prime}_m(\xi_{m;j}-i) \, 
\delta
\xi_{m;j}=0, \quad m=1,2,... \quad j=1,...,n_m,
\end{equation}
where $g(u)=\frac{\pi}{2 \sinh \frac{\pi}{2}u}$ and 
from now on{\footnote{So far we have used the notations of \cite{Arutyunov:2009ax}, where
$s(u)=\frac{g}{4 \, \cosh \frac{\pi}{2}g u}$. }} 
in the rest of the paper $s(u)=\frac{1}{4 \, \cosh \frac{\pi}{2}u}$.
$\delta L_m=(\delta y_m)/(1+y_m)$ is the order $g^8$ perturbation of 
$\log (1+Y_{m|vw})$ in the rescaled variables and $y_m(u)=\lim_{g \to 0} \, 
Y^{o}(\frac{u}{g})$.  
For the twist-two operators the asymptotic form of
$y_m$ corresponds to the $Q(u)=u$ solution of a compact site-$N$ spin~$\frac12$ 
($s=\frac12$) XXX Heisenberg chain with the $N$ inhomogeneity parameters given by the solutions of the
Bethe equations of a site-$2$ non-compact spin minus~$\frac12$ ($s=-\frac12$) Heisenberg chain.

In terms of the solution of (\ref{ADSlin}),(\ref{ADSlinquant}) the wrapping corrected Bethe equations take
the form:
\begin{eqnarray} 
\pi (2n_k+1)&=&J\,
p_k +i\sum_{j=1}^N \log S_{\mathfrak{sl}(2)}^{1_*1_*}(\frac{u_{j}}{g},\frac{u_{k}}{g}) +{\cal
\delta R}_k\,+O(g^9), 
\end{eqnarray}
with ${\cal \delta R}_k$ given by the formula
\begin{equation} \label{delR}
\delta{\cal R}_k=
\delta{\cal R}_k^{(1)}+
\delta{\cal R}_k^{(2)}+
\delta{\cal R}_k^{(3)},
\end{equation}
where $\delta{\cal R}_k^{(1)}$ and $\delta{\cal R}_k^{(3)}$ comes from the small $g$ expansion of the
convolution terms containing the $Y_Q$ functions in (\ref{Y1*}), while $\delta{\cal R}_k^{(2)}$ originates
from the perturbation of the third line and the convolution terms containing the $Y_{1|vw}$ function in 
(\ref{Y1*}). Their explicit form is given by:
\begin{equation} \label{delR1}
\delta{\cal R}_k^{(1)}=
\frac{1}{\pi}\sum_{m=1}^\infty\,\int_{-\infty}^\infty
{\rm d}u\,Y^o_m(u)\,\frac{u-u_k}{(m+1)^2+(u-u_k)^2}\,,
\end{equation}
\begin{equation} \label{delR2}
\delta{\cal R}_k^{(2)}=
\int_{-\infty}^\infty{\rm d}u\,\,\frac{\delta L_1(u)}
{2\sinh\frac{\pi}{2}(u-u_k)}
+\sum_j\frac{\pi\,\delta\xi_{1;j}}{\cosh\frac{\pi}{2}(u_k-\xi_{1;j})}
\end{equation} 
and
\begin{equation} \label{delR3}
\delta{\cal R}_k^{(3)}=
\frac{1}{\pi}\sum_{m=1}^\infty\,\int_{-\infty}^\infty{\rm d}
u\,Y^o_{m+1}(u)\,\left\{{\cal F}_m(u-u_k)-
\frac{u-u_k}{m^2+(u-u_k)^2}\right\},
\end{equation}
where
\begin{equation}
{\cal F}_m(u)=\frac{-i}{4}\left\{\psi\left(\frac{m+iu}{4}\right)
-\psi\left(\frac{m-iu}{4}\right)-\psi\left(\frac{m+2+iu}{4}\right)
+\psi\left(\frac{m+2-iu}{4}\right)\right\}
\end{equation}
with the usual $\psi$ function $\psi(z)=\Gamma^\prime(z)/\Gamma(z)$ and 
no principal value prescription is needed in (\ref{delR2}) since the 
integrand is regular at $u=u_k$. 

The nontrivial part of the calculation is the evaluation of 
$\delta{\cal R}_k^{(2)}$. The details of this calculation are given in 
appendix A. The result is\footnote{Strictly speaking we should use here
the functions shifted away from the real axis by $-i\gamma$ as in appendix~A,
but after the identification (\ref{iden}) we see that potential singularities
along the real axis actually cancel and we correctly get (\ref{BajnokJanik}).}
\begin{equation} \label{resdelR2}
\begin{split}
\delta{\cal R}_k^{(2)}&=\frac{1}{\pi}\,
\sum_{m=1}^{\infty}\int_{-\infty}^{\infty}{\rm d}u\,Y_{m+1}^{o}(u)
\left\{\partial_k\log \hat t_m(u)\right\}\\
&=\frac{1}{\pi}\,
\sum_{m=1}^{\infty}\int_{-\infty}^{\infty}{\rm d}u\,Y_{m+1}^{o}(u)
\left\{\partial_k\log t_m(u)-\frac{r^\prime_m(u-u_k)}{r_m(u-u_k)}\right\}.
\end{split}
\end{equation}
Using
\begin{equation}
\frac{r^\prime_m(x)}{r_m(x)}={\cal F}_m(x)-\frac{2x}{m^2+x^2}
\end{equation}
we find that the transcendental parts cancel and the result can be given
in terms of rational functions:
\begin{equation}
\begin{split} 
\delta{\cal R}_k&=
\frac{1}{\pi}\,\int_{-\infty}^{\infty}{\rm d}u\,Y_1^{o}(u)
\,\frac{u-u_k}{4+(u-u_k)^2}\\
&+\frac{1}{\pi}
\sum_{m=1}^{\infty}\int_{-\infty}^{\infty}{\rm d}u\,Y_{m+1}^{o}(u)
\left\{\partial_k\log t_m(u) 
+\frac{u-u_k}{m^2+(u-u_k)^2}+\frac{u-u_k}{(m+2)^2+(u-u_k)^2}
\right\}\,.
\end{split} 
\end{equation}
This can be compactly written
\begin{equation}
\delta{\cal R}_k=
\frac{1}{2\pi}\,
\sum_{m=1}^{\infty}\int_{-\infty}^{\infty}{\rm d}u\,Y_m^{o}(u)\,
\partial_k\log j_m(u)\,,
\label{finalj}
\end{equation}
where\footnote{The constants $C_2$ and $S_1$ are defined in appendix B.}
\begin{equation}
j_m(u)=\frac{16C_2S_1^2g^8}{(u^2+m^2)^4}\,
\frac{t_{m-1}^2(u)}{
t_0(u-im-i)\,
t_0(u-im+i)\,
t_0(u+im-i)\,
t_0(u+im+i)}\,.
\end{equation}
So far we have not used the explicit form of $Y^o_m(u)$. The crucial 
observation is that
\begin{equation}
Y^o_m(u)=j_m(u)\,,
\label{iden}
\end{equation}
which can be verified using the asymptotic solutions given in \cite{GKV09}.
Thus (\ref{finalj}) finally becomes
\begin{equation}
\delta{\cal R}_k=
\frac{1}{2\pi}\,
\sum_{m=1}^{\infty}\int_{-\infty}^{\infty}{\rm d}u\,
\partial_k\, Y_m^{o}(u)\,,
\label{BajnokJanik}
\end{equation}
which is the Bajnok-Janik formula (\ref{BJv0}).

\section{Conclusions}

In this paper we have shown that the wrapping corrected Bethe equations 
of the twist-two operators of the $sl(2)$ sector coming from the generalized L\"uscher approach
and the TBA equations coincide up to $O(g^8)$. Our considerations are rather
general and are not very sensitive to the details of the state under consideration. The analysis of the
small $g$ behaviour of the terms in the TBA equations suggest that also in the general case, in the calculation 
of
the order $g^{2L}$ correction of the Bethe equations, the linear problem related to the
$Y_{M|vw}$ functions decouple from the rest of the linearized equations and the leading order wrapping 
correction to the Bethe equations is always given by (\ref{delR}), although the positions
of the singularities are different from those of the twist-two operators.
Next it can be recognized that the derivation 
presented in appendix A is insensitive to the distribution of the singularities associated to the
$Y_{M|vw}$ functions indicating that the leading order $g^{2L}$
wrapping correction of the Bethe equations
of any symmetric state (the set of magnon rapidities consists of $\{u_j,-u_j\}$ pairs) 
with R-charge $J$ of the $sl(2)$ sector 
is given by (\ref{BJv0}), with the appropriate asymptotic $Y^{o}_Q$ functions.

\section*{Acknowledgments}

\'A. H. would like to thank Zolt\'an Bajnok
for useful discussions.
This work was supported by the Hungarian
Scientific Research Fund (OTKA) under the grant K 77400.

\appendix

\section{Calculation of $\delta{\cal R}_k^{(2)}$}

As it was mentioned in section 4, 
the asymptotic solution for the Y-system components
associated to $vw$-strings (in the $g\to0$ limit) is identical to a
solution of the rational XXX-model with inhomogeneities.
After suitable rescaling of the independent variable, and introducing
the notation $Y^o_{m\vert vw}\to y_m$, we can recognize that these functions 
satisfy the standard form of the semi-infinite chain of XXX-model 
Y-system equations:
\begin{equation}
y_m(u+i)\,y_m(u-i)=[1+y_{m+1}(u)]\,[1+y_{m-1}(u)],\qquad m=1,2,\dots,
\label{XXX1}
\end{equation}
where $y_0(u)=0$ by convention. The XXX model, its Bethe Ansatz solution
together with the corresponding T-system, Y-system, and Baxter TQ-relations, 
are of course very well known (for a review, see \cite{Wieg}). Here we summarize those elements 
of the solution that we need in our calculation.

A solution of the XXX Y-system relations (\ref{XXX1}) is given in terms of the
solution of the corresponding T-system equations:
\begin{equation}
t_m(u+i)\,t_m(u-i)=t_{m+1}(u)\,t_{m-1}(u)+t_0(u+(m+1)i)\,t_0(u-(m+1)i),
\end{equation}
where all $t_m(u)$ ($m=0,1,\dots$) are polynomials starting with
\begin{equation}
t_0(u)=\prod_{j=1}^N\,(u-v_j).
\end{equation}
Here the set of inhomogeneities $\{v_1,\dots,v_N\}$ can be fixed
arbitrarily. For the AdS/CFT case we are interested in this set has the
special form $\{u_1,-u_1,\dots,u_{N/2},-u_{N/2}\}$, but we will make
this specialization only at the very end of the calculation. 
Starting from a given $t_0$, by solving the Bethe Ansatz equations, we can
build our $t_m$ functions. All these functions are polynomials of degree $N$.
The twist-two states we are studying in this paper correspond to solutions
with a single Bethe root (which is at zero for the symmetric magnon 
distribution $\{u_j,-u_j\}$). These are one of the simplest Bethe Ansatz 
solutions, but their explicit form is not needed in this calculation.
The Y-system functions are given by the well-known formulas
\begin{equation}
y_m(u)=\frac{t_{m+1}(u)\,t_{m-1}(u)}{t_0(u+(m+1)i)\,t_0(u-(m+1)i)}\qquad
m=1,2,\dots
\label{ytt}
\end{equation}
and
\begin{equation}
1+y_m(u)=\frac{t_m(u+i)\,t_m(u-i)}{t_0(u+(m+1)i)\,t_0(u-(m+1)i)}\qquad
m=0,1,\dots
\label{y1tt}
\end{equation}
The position of the roots and poles of $y_m$ and $1+y_m$ are determined
by the roots of the polynomials $t_m$ using these formulas. Actually, in the
TBA framework only those roots that are within the \lq\lq physical strip'', 
i.e. which are real or their distance to the real 
axis is smaller than unity, play any role. We denote by 
$\{\xi_{m;j}\}_{j=1}^{n_m}$ the set of such roots of $t_m$. We define the
sign\footnote{Since later we will shift the integration contour by $-i\gamma$,
real roots are counted as positive. An equivalent way of proceeding would 
have been to shift by $+i\gamma$, in which case the real roots would be 
classified negative.} of a root by
\begin{equation}
\begin{split}
\omega_{m;j}&=+1\qquad\quad{\rm if}
\qquad \,\,\,\,\,\,\, 0\leq{\rm Im}\,\xi_{m;j}<1,\\
\omega_{m;j}&=-1\qquad\quad{\rm if}\qquad -1<{\rm Im}\,\xi_{m;j}<0
\end{split}
\end{equation}
and further define 
\begin{equation}
\tilde\xi_{m;j}=\xi_{m;j}-i\omega_{m;j}.
\end{equation}
We note that $n_0=N$ and $\xi_{0;j}=v_j$,\ $j=1,\dots,N$.
Before proceeding, let us define the T-system elements in a new gauge by
\begin{equation}
\hat t_m(u)=\left\{\prod_{j=1}^N\,r_m(u-v_j)\right\}\,t_m(u)\,,
\end{equation}
where
\begin{equation}
r_m(u)=\frac{1}{4}\,\frac{\gamma(2+m+iu)\,\gamma(2+m-iu)}
{\gamma(4+m+iu)\,\gamma(4+m-iu)}
\end{equation}
with $\gamma(u)=\Gamma(u/4)$. It is easy to verify that 
(\ref{ytt}-\ref{y1tt}) are simplified to
\begin{equation}
y_m(u)=\hat t_{m+1}(u)\,\hat t_{m-1}(u)\qquad
m=1,2,\dots,
\end{equation}
\begin{equation}
1+y_m(u)=\hat t_m(u+i)\,\hat t_m(u-i)\qquad
m=0,1,\dots
\label{y1hatthatt}
\end{equation}
and that $t_m$ and $\hat t_m$ have the same physical roots 
(and neither have poles).

To avoid any singularities, as explained in the main text we shift the real line by a small
amount $-i\gamma$ in the negative imaginary direction. This means that
the new physical strip becomes
\begin{equation}
-1-\gamma<{\rm Im}\,u<1-\gamma\,,
\end{equation}
which explains why real roots are classified here as positive. (We have to 
choose $\gamma$ small enough so that no physical root is lost or no new
physical root is created by this shift. This is possible if there are no roots
on the boundary of the original physical strip.)

It is now standard to translate the functional relation (\ref{y1hatthatt})
into a TBA type integral equation\footnote{Using the notation 
$f^{-\gamma}(u)=f(u-i\gamma)$ for any function $f$.}
\begin{equation}
\hat t^{-\gamma}_m(u)=\tau_m(u)\,\exp\left\{(s\star L_m)(u)\right\}\,,
\end{equation}
where $L_m(u)=\log(1+y^{-\gamma}_m(u))$ and
\begin{equation}
\tau_m(u)=\prod_{j=1}^{n_m}\,\tanh\frac{\pi}{4}(u-i\gamma-\xi_{m;j}).
\end{equation}
Similarly the Y-system equations can be transformed to the TBA equations
\begin{equation}
y^{-\gamma}_m(u)=\tau_{m+1}(u)\,\tau_{m-1}(u)\,
\exp\left\{(s\star [L_{m+1}+L_{m-1}])(u)\right\}\,,\qquad m=1,2,\dots
\label{XXXTBA}
\end{equation}
These are supplemented by the quantization conditions
\begin{equation}
y_m(\tilde \xi_{m;j})=-1\,,\qquad m=1,2,\dots,\qquad j=1,\dots,n_m\,,
\label{XXXQC}
\end{equation}
which follow from (\ref{y1hatthatt}).

We now \lq\lq linearize'' the TBA equations (\ref{XXXTBA}) by taking their
logarithmic derivative with respect to one of the inhomogeneity parameters,
$v_k$:
\begin{equation}
\partial_k\,\ell_m=-H_{m+1}-H_{m-1}+s\star\left(\partial_kL_{m+1}+\partial_k
L_{m-1}\right)\,,\qquad m=1,2,\dots,
\label{derTBA}
\end{equation}
where
\begin{equation}
\partial_k=\frac{\partial}{\partial v_k}\,,\qquad
\ell_m(u)=\log y^{-\gamma}_m(u),\qquad H_{m}(u)=\sum_{j=1}^{n_m}\,
Q_{m;j}(u)\,\partial_k\xi_{m;j}
\end{equation}
with
\begin{equation}
Q_{m;j}(u)=g(u-i\gamma-\xi_{m;j})\,,\qquad 
g(u)=\frac{\pi}{2\sinh\frac{\pi}{2}u}\,.
\end{equation}
Similarly the linearized form of the quantization conditions is 
\begin{equation}
\partial_k\ell_m(\tilde\xi_{m;j}+i\gamma)-y^\prime_m(\tilde\xi_{m;j})\,
\partial_k\xi_{m;j}=0\,,\qquad m=1,2,\dots,\qquad j=1,\dots,n_m\,.
\end{equation}

The above linearized problem is very similar to 
(\ref{ADSlin})-(\ref{ADSlinquant}),
obtained by the linearization of the full AdS/CFT TBA system in the 
main text (around the same XXX-model Bethe Ansatz solution). 
Rewriting those equations using
the definitions introduced above and shifting the contour by $-i\gamma$
we get
\begin{equation}
\delta\,\ell_m=-h_{m+1}-h_{m-1}+s\star\left(\delta L_{m+1}+\delta
L_{m-1}\right)+i_m\,,\qquad m=1,2,\dots,
\label{linTBA}
\end{equation}
where
\begin{equation}
h_{m}(u)=\sum_{j=1}^{n_m}\,
Q_{m;j}(u)\,\delta\xi_{m;j}\,,\qquad
i_m=-s\star X_m\,,\qquad X_m(u)=Y^o_{m+1}(u-i\gamma)\,.
\label{linsource}
\end{equation}
We note that by convention $h_0=\delta L_0=0$ here.
After the shift the linearized quantization conditions become
\begin{equation}
\delta\ell_m(\tilde\xi_{m;j}+i\gamma)-y^\prime_m(\tilde\xi_{m;j})\,
\delta\xi_{m;j}=0\,,\qquad m=1,2,\dots,\qquad j=1,\dots,n_m\,.
\end{equation}

Apart from the fact that the deviation from the given Bethe Ansatz solution
is caused by changing one of the inhomogeneity parameters in the first case 
and coupling to other nodes of the AdS/CFT diagram in the second, the only 
difference between (\ref{derTBA}) and (\ref{linTBA}) is that in the former
there are no source terms and $H_0(u)=g(u-i\gamma-v_k)\not=0$. We can,
however, change our conventions by putting $H_0=0$ also in this case and
compensating this by adding a source term
\begin{equation}
i_m(u)=-\delta_{m1}\,g(u-i\gamma-v_k)\,.
\label{dersource}
\end{equation}
After these changes the two linear problems have identical structure.

Let us now write out this linear structure in some detail. Arranging the 
two types of unknowns as two (infinite component) column vectors
\begin{equation}
\delta L=
\begin{pmatrix}
\delta L_1(u)\\
\delta L_2(u)\\
\vdots
\end{pmatrix}
\qquad\quad
\delta\xi=
\begin{pmatrix}
\delta \xi_{1;j}\\
\delta \xi_{2;j}\\
\vdots
\end{pmatrix}
\end{equation}
we can then write the linearized TBA equations schematically as
\begin{equation}
M_{11}\,\delta L+M_{12}\,\delta\xi=I_1
\end{equation}
and the linearized quantization conditions as
\begin{equation}
M_{21}\,\delta L+M_{22}\,\delta\xi=I_2.
\end{equation}
Here
\begin{equation}
I_1=
\begin{pmatrix}
i_1(u)\\
i_2(u)\\
\vdots
\end{pmatrix}
\end{equation}
and, since for later convenience we multiply the quantization conditions
by $2\pi i\omega_{m;j}$,
\begin{equation}
I_2=
\begin{pmatrix}
-2\pi i\omega_{1;j}\,i_1(\tilde\xi_{1;j}+i\gamma)\\
-2\pi i\omega_{2;j}\,i_2(\tilde\xi_{2;j}+i\gamma)\\
\vdots
\end{pmatrix}\,.
\end{equation}
The operator matrices $M_{11}$ etc. are as follows.
\begin{equation}
M_{11}=\begin{pmatrix}
D_1&-\sigma&0&0&\dots\\
-\sigma&D_2&-\sigma&0&\dots\\
0&-\sigma&D_3&-\sigma&\dots\\
0&0&-\sigma&D_4&\dots\\
&&\vdots&&
\end{pmatrix}\,,
\end{equation}
where $D_m=1+1/y_m$, $\sigma=s\star$,
\begin{equation}
M_{12}=\begin{pmatrix}
0&V_2&0&0&\dots\\
V_1&0&V_3&0&\dots\\
0&V_2&0&V_4&\dots\\
0&0&V_3&0&\dots\\
&&\vdots&&
\end{pmatrix}\,,
\end{equation}
with $V_{m;j}(u)=Q_{m;j}(u)$,
\begin{equation}
M_{21}=\begin{pmatrix}
0&V_1^T&0&0&\dots\\
V_2^T&0&V_2^T&0&\dots\\
0&V_3^T&0&V_3^T&\dots\\
0&0&V_4^T&0&\dots\\
&&\vdots&&
\end{pmatrix}\,,
\end{equation}
where $^T$ denotes transposition.
Finally the matrix elements of $M_{22}$ are given by the formula
\begin{equation}
M_{22mm^\prime;jj^\prime}=-2\pi i\omega_{m;j}\,y^\prime_m(\tilde\xi_{m;j})\,
\delta_{mm^\prime}\,\delta_{jj^\prime}+
(2\pi)^2\,s(\xi_{m;j}-\xi_{m^\prime; j^\prime})\,(\delta_{m+1\,m^\prime}+
\delta_{m-1\,m^\prime})\,.
\end{equation}
The crucial observation is that
\begin{equation}
M_{11}^T=M_{11}\,,\qquad M_{12}^T=M_{21}\,,\qquad M_{22}^T=M_{22}\,,
\end{equation}
and consequently the big operator matrix of the linear problem is symmetric:
\begin{equation}
{\bf M}^T={\bf M}\,,\qquad\quad 
{\bf M}=\begin{pmatrix}M_{11}&M_{12}\\M_{21}&M_{22}\end{pmatrix}\,.
\end{equation}
This means that, if the inverse operator ${\bf R}$ exists (which we assume)
then it must also be symmetric:
\begin{equation}
{\bf R}^T={\bf R}\,,\qquad {\bf R}={\bf M}^{-1}\,.
\end{equation}
Writing it as a hypermatrix
\begin{equation}
{\bf R}=\begin{pmatrix}A&B\\C&D\end{pmatrix}
\end{equation}
this symmetry property, in terms of its components, reads:
\begin{equation}
A_{m,m^\prime}(u,v)=A_{m^\prime, m}(v,u)\,,\quad
B_{m,m^\prime;j}(u)=C_{m^\prime;j,m}(u)\,,\quad
D_{m;j,m^\prime;j^\prime}=D_{m^\prime;j^\prime,m;j}
\label{sym}
\end{equation}
Using the components of the inverse matrix and the source term 
(\ref{dersource}), we can write
\begin{equation}
\begin{split}
\partial_kL_m(u)&=\int_{-\infty}^\infty{\rm d}v\,g(v_k+i\gamma-v)\,
A_{1,m}(v,u)-(2\pi)^2\,\sum_j C_{1;j,m}(u)\,s(v_k-\xi_{1;j})\,,\\
\partial_k\xi_{m;j}&=\int_{-\infty}^\infty{\rm d}v\,g(v_k+i\gamma-v)\,
B_{1,m;j}(v)-(2\pi)^2\,\sum_{j^\prime}\,
s(v_k-\xi_{1;j^\prime})\,D_{1;j^\prime,m;j}\,,
\end{split}
\label{dersol}
\end{equation}
where we already used the symmetry properties (\ref{sym}).

For the AdS/CFT case, the source terms are of the form (\ref{linsource})
and the solution of the linear problem will depend linearly on the
functions $X_m$. We define:
\begin{equation}
\begin{split}
\delta L_m(u)&=\sum_{m^\prime}\,\int_{-\infty}^\infty\,{\rm d}w\,
\delta L_m^{(m^\prime)}(u,w)\,X_{m^\prime}(w)\,,\\
\delta \xi_{m;j}&=\sum_{m^\prime}\,\int_{-\infty}^\infty\,{\rm d}w\,
\delta \xi_{m;j}^{(m^\prime)}(w)\,X_{m^\prime}(w)\,.
\end{split}
\end{equation}
To calculate the L\"uscher correction, we only need $\delta L_1$ and
$\delta\xi_{1;j}$:
\begin{equation}
\delta{\cal R}_k^{(2)}=\frac{1}{\pi}
\int_{-\infty}^\infty{\rm d}v\,g(v-i\gamma-v_k)\,\delta L_1(v)+
4\pi\sum_j s(v_k-\xi_{1;j})\delta\xi_{1;j}
\label{lus}
\end{equation}
and we can define
\begin{equation}
\delta{\cal R}_k^{(2)}=
\sum_m\int_{-\infty}^\infty{\rm d}w\,\delta{\cal R}_k^{(2)(m)}(w)
X_m(w)\,.
\end{equation}
Using the inverse operator we can write
\begin{equation}
\begin{split}
\delta L_1^{(m)}(u,w)&=-\int_{\infty}^\infty{\rm d}v\,
A_{1,m}(u,v)\,s(v-w)+\sum_jB_{1,m;j}(u)Q_{m;j}(w)\,,\\
\delta \xi_{1;j}^{(m)}(w)&=-\int_{\infty}^\infty{\rm d}v\,
C_{1;j,m}(v)\,s(v-w)+\sum_{j^\prime}D_{1;j,m;j^\prime}Q_{m;j^\prime}(w)\,.
\end{split}
\end{equation}
Note that exactly the same matrix elements of the inverse operator ${\bf R}$
appear here as in (\ref{dersol}). Substituting these formulas into (\ref{lus})
and using the relations (\ref{dersol}) we get
\begin{equation}
\begin{split}
\pi \, \delta{\cal R}_k^{(2)(m)}(w)&=\int_{\infty}^\infty{\rm d}v\,
\partial_kL_m(v)\,s(v-w)-\sum_j\left(\partial_k\xi_{m;j}\right)\,Q_{m;j}(w)\\
&=\partial_k(s\star L_m)(w)+\partial_k\log\tau_m(w)
=\partial_k\log \hat t^{-\gamma}_m(w)\,.
\end{split}
\end{equation}
Finally we can write
\begin{equation}
\begin{split}
\delta{\cal R}_k^{(2)}&=\frac{1}{\pi}
\sum_{m=1}^\infty\int_{-\infty}^\infty{\rm d}w\,
X_m(w)\partial_k\log \hat t^{-\gamma}_m(w)\\
&=\frac{1}{\pi}\sum_{m=1}^\infty\int_{-\infty}^\infty{\rm d}w\,
Y^o_{m+1}(w-i\gamma)\left\{-\frac{r_m^\prime(w-i\gamma-v_k)}
{r_m(w-i\gamma-v_k)}+
\partial_k\log t^{-\gamma}_m(w)\right\}\,.
\end{split}
\end{equation}
We can shift the integration contour back to the real axis at the end of the
calculation.

\section{Simplification of the Bajnok-Janik formula}

This appendix is based on the results of ref. \cite{Bajnok:2008qj}. 
Unfortunately our conventions are different from that of this 
paper\footnote{ $g=2g^{\rm BJL}$, $u_k=2u_k^{\rm BJL}$}. Here we write
all formulae in our conventions.

The L\"uscher correction to the energy is given by
\begin{equation}
\Delta E=-\frac{1}{2\pi}\sum_{Q=1}^\infty\int_{-\infty}^\infty\,
{\rm d}q\,\frac{g^4}{(q^2+Q^2)^2}\varepsilon_Q\,,
\end{equation}
where
\begin{equation}
\varepsilon_Q={\rm STr}\left\{S_{Q;1}(q,v_1)S_{Q;1}(q,v_2)\dots
S_{Q;1}(q,v_N)\right\}.
\end{equation}
Here $v_1=u_1$, $v_2=-u_1$, etc. is the symmetric magnon configuration. We
want to calculate $\varepsilon_Q$ in the lowest nontrivial order.

Similarly the correction to the Bethe-Yang equations is given by
\begin{equation}
\delta {\cal R}_1=-\frac{1}{2\pi}\sum_{Q=1}^\infty\int_{-\infty}^\infty\,
{\rm d}q\,\frac{g^4}{(q^2+Q^2)^2}\Phi_Q\,,
\end{equation}
where
\begin{equation}
\Phi_Q={\rm STr}\left\{S^\prime_{Q;1}(q,v_1)S_{Q;1}(q,v_2)\dots
S_{Q;1}(q,v_N)\right\}.
\end{equation}
Here $^\prime$ means derivative withe respect to the variable $q$ and again,
we are interested in the lowest non-trivial order in $g^2$.

In what follows we concentrate on the contributions coming from a fixed
$Q$ sector. $S_{Q;1}(q,u)$, the S-matrix in this sector is a product of a
scalar factor $\sigma(q,u)$ and a tensor product of two identical matrix
factors. These matrices can be diagonalized in a $u$-independent way and
can be written as
\begin{equation}
G(q){\cal D}(q,u)G^{-1}(q)\,,
\end{equation}
where ${\cal D}$ is diagonal:
\begin{equation}
{\cal D}(q,u)=\langle S^\alpha_j(q,u)\rangle\,,
\end{equation}
and the eigenvalues can be grouped in such a way that their expansion is of
the form
\begin{equation}
S^\alpha_j(q,u)=K^\alpha(u)\,A_j(q,u)\,\left\{1+g^2\delta^\alpha_j(q,u)\right
\}+{\rm O}(g^4)\,,
\end{equation}
where $j=0,1,\dots,Q-1$, $\alpha=0,1,2,3$ with 0,2 bosons, 1,3 fermions
and since
\begin{equation}
K^\alpha(u)K^\alpha(-u)=1\,,
\end{equation}
$K^\alpha(u)$ plays no role for our symmetric configuration.

The scalar factor is given explicitly by
\begin{equation}
\sigma(q,u)=\frac{(u+i)^2}{\{(q-u)^2+(Q+1)^2\}\,\{(q-u)^2+(Q-1)^2\}}
\end{equation}
and
\begin{equation}
A_j(q,u)=q-u+i(1+2j-Q)\,.
\end{equation}

Using these building blocks, we can write
\begin{equation}
\varepsilon_Q=\sigma_1\sigma_2\dots\sigma_N\,m^2\,,
\end{equation}
where $\sigma_k=\sigma(q,v_k)$ and the matrix part is
\begin{equation}
m=\sum_{j,\alpha}(-1)^\alpha S^\alpha_{1j}S^\alpha_{2j}\dots S^\alpha_{Nj}\,,
\end{equation}
where $S^\alpha_{kj}=S^\alpha_j(q,v_k)$.

The crucial observation of ref. \cite{Bajnok:2008qj} is that
\begin{equation}
\sum_\alpha (-1)^\alpha=0\,,\qquad\quad
\sum_\alpha(-1)^\alpha \delta^\alpha_j(q,u)=ih(u)\,r_j(q)\,,
\end{equation}
where
\begin{equation}
h(u)=\frac{2}{1+u^2}\,,\qquad\quad
r_j(q)=\frac{1}{q+i(2j-Q)}-\frac{1}{q+i(2+2j-Q)}\,.
\end{equation}
This is why the leading term vanishes and the next term becomes simple:
\begin{equation}
m=iC_1g^2\sum_{j=0}^{Q-1}A_{1j}A_{2j}\dots A_{Nj}r_j(q)\,,
\end{equation}
where $A_{kj}=A_j(q,v_k)$ and
\begin{equation}
C_1=\sum_{k=1}^N h(v_k)=2\sum_{k=1}^{N/2}h(u_k)=2S_1(N)
=2\sum_{i=1}^N\frac{1}{i}\,.
\end{equation}
In what follows, $C_1$, together with a similar constant coming from
the scalar factors,
\begin{equation}
C_2(N)=\prod_{k=1}^N(1+v_k^2)=\left(\frac{2^N(N!)^3}{(2N)!}\right)^2
\label{C2}
\end{equation}
are treated as numerical constants\footnote{The value (\ref{C2}) can be 
calculated using the generalized hypergeometric representation of the 
leading order particle rapidities \cite{Bajnok:2008qj}.}, which take definite
numerical value for our twist-two states.

$\Phi_Q$ can be built from the same building blocks:
\begin{equation}
\Phi_Q=\sigma^\prime_1\sigma_2\dots\sigma_N\,m^2+
2\sigma_1\sigma_2\dots\sigma_N\,m\tilde m\,,
\end{equation}
where
\begin{equation}
\tilde m=iC_1g^2\sum_{j=0}^{Q-1}A^\prime_{1j}A_{2j}\dots A_{Nj}r_j(q)+
ig^2h(v_1)\sum_{j=0}^{Q-1}A_{1j}A_{2j}\dots A_{Nj}r^\prime_j(q)\,.
\label{tildem}
\end{equation}
This is obtained from the derivative acting on the eigenvalues. A potential
additional term, where the derivative is acting on the diagonalizing matrices
is of the form
\begin{equation}
{\rm STr}\left\{\left[G^{-1}G^\prime,{\cal D}\right]({\rm diag})\right\}
\end{equation}
and since the commutator term is off-diagonal and the rest diagonal, this
vanishes. The second term in (\ref{tildem}) can be dropped, because the 
summands are odd under the symmetry $j\rightarrow Q-1-j$, $q\rightarrow -q$
and thus do not contribute after summation and integration, multiplied by
a $q$-even function.

All the terms that are left depend only on the difference $q-u$ (like in 
relativistic models) and therefore the derivative with respect to $q$ can be
replaced by the derivative with respect to the first rapidity $u_1$. This
leads to the identity
\begin{equation}
\Phi_Q\approx-\frac{1}{2}\frac{\partial}{\partial u_1}\varepsilon_Q\,.
\label{Phiidentity}
\end{equation}
$\approx$ here means that the equality holds only after integrating both
sides, multiplied by even functions. To prove (\ref{Phiidentity}) we can
use a little lemma stating that for $f$, $g$ even functions,
\begin{equation}
\int_{-\infty}^\infty{\rm d}qf^\prime(q-\alpha)f(q+\alpha)g(q)=
-\frac{1}{2}\frac{\partial}{\partial\alpha}
\int_{-\infty}^\infty{\rm d}qf(q-\alpha)f(q+\alpha)g(q)\,.
\end{equation}
This applies directly to the $\sigma^\prime_1\sigma_2\dots\sigma_N$ 
term and for the first term in (\ref{tildem}) we can use it combined with the
$j\rightarrow Q-1-j$ symmetry.

Let us compute $\varepsilon_Q$ explicitly. We find
\begin{equation}
\varepsilon_Q=\frac{16g^4C_2(N)S_1^2(N)t^2_{Q-1}(q)}
{(q^2+Q^2)^2t_0(q+i+iQ)t_0(q-i-iQ)t_0(q-i+iQ)t_0(q+i-iQ)}\,,
\end{equation}
where
\begin{equation}
t_0(q)=\prod_{k=1}^N(q-v_k)
\label{inhom}
\end{equation}
and $t_m(q)$ are the XXX model T-system elements corresponding to the
inhomogeneities (\ref{inhom}) and Baxter's $Q$ function
\begin{equation}
Q^{\rm Baxter}(q)=q\,,
\end{equation}
i.e. one Bethe root at zero. 

Thus our final result is
\begin{equation}
\delta{\cal R}_1=
\frac{1}{4\pi}\frac{\partial}{\partial u_1}
\sum_{Q=1}^\infty\int_{-\infty}^\infty
{\rm d}q\,Y^o_Q(q)
\label{finres}
\end{equation}
with
\begin{equation}
Y^o_Q(q)=\frac{16g^8C_2(N)S_1^2(N)t^2_{Q-1}(q)}
{(q^2+Q^2)^4t_0(q+i+iQ)t_0(q-i-iQ)t_0(q-i+iQ)t_0(q+i-iQ)}\,.
\end{equation}
It is not difficult to see that in (\ref{finres}) we can make the substitution
\begin{equation}
\frac{\partial}{\partial u_1}
\rightarrow2\frac{\partial}{\partial v_1}=2\partial_1\,,
\end{equation}
where the derivative is understood as follows. First $v_1,v_2,\dots,v_N$
are treated as independent inhomogeneity parameters, and only after the
integration the derivative with respect to $v_1$ was taken is the configuration
restricted again to the symmetric one.

\section{The Bajnok-Janik formula in relativistic models}

The generalized L\"uscher approach is also valid in relativistic integrable
models, like the Sine-Gordon model, nonlinear $\sigma$-models and other related
models. In this appendix we summarize the results for the O$(2)$, O$(3)$ and 
O$(4)$ nonlinear $\sigma$-models and the SU$(n)$ principal model. 
We have proven the formulas using the known TBA/NLIE description of these 
models. The details of the derivation will be published elsewhere.

In the simple cases corresponding to the O$(n)$ $\sigma$-models for $n=2,3$
and 4 there is only one type of particles of mass $\mu$ and there are no 
bound states. For a state consisting of $r$ particles of rapidities
$\{\theta_1,\dots,\theta_r\}$ in a very large periodic box of length $L$
the energy of the system is simply
\begin{equation}
E^{(0)}=\sum_{j=1}^r\,\mu\cosh\theta_j
\end{equation}
and the particle rapidities are subject to the quantization conditions
\begin{equation}
QC_k^{(0)}(\theta_1,\dots,\theta_r)
={\rm e}^{i\mu L\sinh\theta_k}\,{\rm e}^{i{\cal R}_k}=-1\,,
\qquad\quad k=1,\dots,r\,,
\end{equation}
where
\begin{equation}
{\rm e}^{i{\cal R}_k}=\sigma(\theta_k\vert\theta_1,\dots,\theta_r)
\end{equation}
and $\sigma(\theta\vert\theta_1,\dots,\theta_r)$ is the eigenvalue of the
transfer matrix (constructed from the unitary and crossing symmetric 
physical S-matrix) corresponding to the given state.

The energy expression that includes the exponentially small first correction 
to the energy is given by L\"uscher's formula:
\begin{equation}
E^{(1)}=E^{(0)}+\delta E=E^{(0)}-
\frac{1}{2\pi}\,\int_{-\infty}^\infty\,{\rm d}\theta\,\mu\cosh\theta\,
{\rm e}^{-\mu L\cosh\theta}\,\sigma\left(\theta+\frac{i\pi}{2}
\Big\vert\theta_1,
\dots,\theta_r\right)\,.
\end{equation}
At the same exponential order the quantization condition is modified to
\begin{equation}
QC_k^{(1)}(\theta_1,\dots,\theta_r)=
QC_k^{(0)}(\theta_1,\dots,\theta_r)\left\{1+i\,\delta {\cal R}_k\right\}
=-1\,,\qquad\quad k=1,\dots,r\,,
\end{equation}
where
\begin{equation}
\delta {\cal R}_k(\theta_1,\dots,\theta_r)=
\frac{1}{2\pi}\,\int_{-\infty}^\infty\,{\rm d}\theta\,
{\rm e}^{-\mu L\cosh\theta}\,\partial_k\,
\sigma\left(\theta+\frac{i\pi}{2}\Big\vert\theta_1,
\dots,\theta_r\right)
\end{equation}
with $\partial_k=\partial/\partial\theta_k$.

For the SU$(n)$ principal model the formulae are more complicated since
in this model there are several types of particles, one corresponding to
each fundamental representation \cite{Hollo}. 
They can be indexed from 1 to $n-1$, 1 corresponding to the defining 
(vector) representation and $n-1$ to their antiparticles. 
We denote the corresponding masses by $\mu_a$ and
by $\sigma^a(\theta\vert\theta_1,\dots,\theta_r)$ the transfer matrix 
eigenvalues corresponding to the $a^{\rm th}$ fundamental representation
in the auxiliary space. For simplicity we here consider states with all 
particles belonging to the $n-1$ (anti-vector) representation only.
In this case the L\"uscher correction to the energy is given by
\begin{equation}
\begin{split}
E^{(1)}&=E^{(0)}+\delta E=\sum_{j=1}^r\,\mu_{n-1}\cosh\theta_j\\
&+\frac{i}{2\pi}\,\sum_{a=1}^{n-1}\,\mu_a\,
\int_{-\infty}^\infty\,{\rm d}\theta\,\sinh\left(\theta+\frac{i\pi}{n}\right)
\, {\rm e}^{i\mu_a L\sinh\left(\theta+\frac{i\pi}{n}\right)}\,
\sigma^a\left(\theta+\frac{i\pi}{n}\Big\vert\theta_1,
\dots,\theta_r\right)\,.
\end{split}
\end{equation}
Similarly the quantization conditions for $k=1,\dots,r$ can be written as
\begin{equation}
QC_k^{(1)}=QC_k^{(0)}\left\{1+i\,\delta{\cal R}_k\right\}=
{\rm e}^{i\mu_{n-1}L\sinh\theta_k}\,\sigma^{n-1}(\theta_k\vert
\theta_1,\dots,\theta_r)\,\left\{1+i\,\delta{\cal R}_k\right\}=-1\,,
\end{equation}
where
\begin{equation}
\delta{\cal R}_k(\theta_1,\dots,\theta_r)=
\frac{1}{2\pi}\,\sum_{a=1}^{n-1}\,
\int_{-\infty}^\infty\,{\rm d}\theta\,
\, {\rm e}^{i\mu_a L\sinh\left(\theta+\frac{i\pi}{n}\right)}\,
\partial_k\,\sigma^a\left(\theta+\frac{i\pi}{n}\Big\vert\theta_1,
\dots,\theta_r\right)\,.
\end{equation}



\begin{thebibliography}{999}

\bibitem{adscft} J.~M.~Maldacena,
``The large N limit of superconformal field theories and supergravity,''
Adv.\ Theor.\ Math.\ Phys.\  {\bf 2} (1998) 231
[Int.\ J.\ Theor.\ Phys.\  {\bf 38} (1999) 1113], [hep-th/9711200];\\
S.~S.~Gubser, I.~R.~Klebanov and A.~M.~Polyakov,
``Gauge theory correlators from non-critical string theory,''
Phys.\ Lett.\ B {\bf 428} (1998) 105, [hep-th/9802109];\\
E.~Witten,
``Anti-de Sitter space and holography,''
Adv.\ Theor.\ Math.\ Phys.\  {\bf 2} (1998) 253, [hep-th/9802150].

\bibitem{BS}
  N.~Beisert and M.~Staudacher,
  ``Long-range PSU(2,2$|$4) Bethe ansaetze for gauge theory and strings,''
  Nucl.\ Phys.\ B {\bf 727}, 1 (2005)
  [hep-th/0504190].

\bibitem{AJK}
  J.~Ambjorn, R.~A.~Janik and C.~Kristjansen,
  ``Wrapping interactions and a new source of corrections to the spin-chain / string duality,''
 {\slshape   Nucl.\ Phys.\  B }{\bf 736} (2006) 288
  [arXiv:hep-th/0510171].

\bibitem{BJ08}
  Z.~Bajnok and R.~A.~Janik,
  ``Four-loop perturbative Konishi from strings and finite size effects for multiparticle states,''
 {\slshape   Nucl.\ Phys.\  B }{\bf 807} (2009) 625
  [arXiv:0807.0399 [hep-th]].

\bibitem{JL07}
  R.~A.~Janik and T.~Lukowski,
  ``Wrapping interactions at strong coupling -- the giant magnon,''
 {\slshape   Phys.\ Rev.\  D }{\bf 76} (2007) 126008
  [arXiv:0708.2208 [hep-th]].

\bibitem{Luscher85}
  M.~Luscher,
  ``Volume Dependence of the Energy Spectrum in Massive Quantum Field Theories. 1. Stable Particle States,''
 {\slshape   Commun.\ Math.\ Phys.\  }{\bf 104} (1986) 177.

\bibitem{Sieg}
  F.~Fiamberti, A.~Santambrogio, C.~Sieg and D.~Zanon,
  ``Wrapping at four loops in N=4 SYM,''
  Phys.\ Lett.\  B {\bf 666} (2008) 100
  [arXiv:0712.3522 [hep-th]].

\bibitem{Vel}
  V.~N.~Velizhanin,
  ``The Four-Loop Konishi in N=4 SYM,''
  arXiv:0808.3832 [hep-th].

\bibitem{Bajnok:2008qj}
  Z.~Bajnok, R.~A.~Janik and T.~Lukowski,
  ``Four loop twist two, BFKL, wrapping and strings,''
  Nucl.\ Phys.\  B {\bf 816} (2009) 376
  [arXiv:0811.4448 [hep-th]].

\bibitem{KL02}
A.~V.~Kotikov and L.~N.~Lipatov,
``DGLAP and BFKL Evolution Equations in the ${\mathcal{N}}\!=4$ Supersymmetric Gauge   Theory,''
Nucl.\ Phys.\  B {\bf 661} (2003) 19
[Erratum-ibid.\  B {\bf 685} (2004) 405]
[arXiv:hep-ph/0208220].

\bibitem{40_5}
  A.~V.~Kotikov, L.~N.~Lipatov, A.~Rej, M.~Staudacher and V.~N.~Velizhanin,
  ``Dressing and Wrapping,''
  J.\ Stat.\ Mech.\  {\bf 0710} (2007) P10003
  [arXiv:0704.3586 [hep-th]].

\bibitem{BJ09}
  Z.~Bajnok, A.~Hegedus, R.~A.~Janik and T.~Lukowski,
  ``Five loop Konishi from AdS/CFT,''
  Nucl. Phys. B {\bf 827} (2010), 426-456,
  arXiv:0906.4062 [hep-th].

\bibitem{Lukowski:2009ce}
  T.~Lukowski, A.~Rej and V.~N.~Velizhanin,
  ``Five-Loop Anomalous Dimension of Twist-Two Operators,''
  Nucl.Phys.B {\bf 831}   (2010), 105-132,
  arXiv:0912.1624 [hep-th].

\bibitem{AF07}
  G.~Arutyunov and S.~Frolov,
  ``On String S-matrix, Bound States and TBA,''
  JHEP {\bf 0712} (2007) 024, hep-th/0710.1568.

\bibitem{AF09a}
  G.~Arutyunov and S.~Frolov,
  ``String hypothesis for the $AdS$ mirror,''
  JHEP {\bf 0903} (2009) 152
  [arXiv:0901.1417 [hep-th]].

\bibitem{AF09b}
  G.~Arutyunov and S.~Frolov,
  ``Thermodynamic Bethe Ansatz for the $AdS$ Mirror Model,''
  JHEP {\bf 0905} (2009) 068
  [arXiv:0903.0141 [hep-th]].

\bibitem{AF09d}
  G.~Arutyunov and S.~Frolov,
  ``Simplified TBA equations of the $AdS_5 \times S^5$ mirror model,''
  arXiv:0907.2647 [hep-th].

\bibitem{Bombardelli:2009ns}
  D.~Bombardelli, D.~Fioravanti and R.~Tateo,
  ``Thermodynamic Bethe Ansatz for planar AdS/CFT: a proposal,''
  J.\ Phys.\ A  {\bf 42} (2009) 375401
  [arXiv:0902.3930].

\bibitem{GKKV09}
  N.~Gromov, V.~Kazakov, A.~Kozak and P.~Vieira,
  ``Integrability for the Full Spectrum of Planar AdS/CFT II,''
  arXiv:0902.4458v3 [hep-th].

\bibitem{DT}
P. Dorey, R. Tateo, �Excited states by analytic continuation of TBA equations,� Nucl.
Phys. B 482, 639 (1996) [arXiv:hep-th/9607167]. 

\bibitem{GKV09b}
  N.~Gromov, V.~Kazakov and P.~Vieira,
  ``Exact AdS/CFT spectrum: Konishi dimension at any coupling,''
  arXiv:0906.4240 [hep-th].

\bibitem{Arutyunov:2009ax}
  G.~Arutyunov, S.~Frolov and R.~Suzuki,
  ``Exploring the mirror TBA,''
  arXiv:0911.2224 [hep-th].

\bibitem{Gromov}
  N.~Gromov,
  ``Y-system and Quasi-Classical Strings,''
  arXiv:0910.3608 [hep-th].

\bibitem{ujKazi}
N. Gromov, V. Kazakov, Z. Tsuboi,
"$PSU(2,2|4)$ character of quasiclasical AdS/CFT"
arXiv:1002.3981 [hep-th]


\bibitem{AFS}
G.~Arutyunov, S.~Frolov and R.~Suzuki,
``Five-loop Konishi from the Mirror TBA'',
arXiv:1002.1711 [hep-th].

\bibitem{BHL}
J. Balog, \'A. Heged\H{u}s,
"5-loop Konishi from linearized TBA and the XXX magnet"
arXiv:1002.4142 [hep-th]


\bibitem{GKV09}
  N.~Gromov, V.~Kazakov and P.~Vieira,
  ``Integrability for the Full Spectrum of Planar AdS/CFT,''
  arXiv:0901.3753 [hep-th].



\bibitem{SCFT1}
 M. Staudacher, 
"The factorized S-matrix of CFT/AdS," JHEP 0505 (2005) 054,
[ar:Xiv:hep-th/0412188].


\bibitem{Wieg}
P.~Wiegmann,
``Bethe Ansatz and Classical Hirota Equation,''
  Int.\ J.\ Mod.\ Phys.\  B {\bf 11} (1997) 75
  [arXiv:cond-mat/9610132].

\bibitem{Hollo}
  T.~J.~Hollowood,
 ``From A(m-1) trigonometric S matrices to the thermodynamic Bethe ansatz,''
  Phys.\ Lett.\  B {\bf 320} (1994) 43
  [arXiv:hep-th/9308147].



\end{thebibliography}
\end{document}